\documentstyle[12pt,epsf]{article}
\newcommand{\nobody}{\rule{0ex}{1ex}}
\newtheorem{fig}[figure]{Figure}
\begin{document} 
\begin{center}
{\LARGE\bf Constraints on Couplings\\ of Extra Gauge Bosons
 from $e^+e^-\rightarrow \nu\bar{\nu}\gamma$}
\vspace{1cm}\\
Stephen  Godfrey$^a$, Pat Kalyniak$^a$, Basim Kamal$^a$
and Arnd Leike$^b,$\\
$\nobody^a$ {\it
Phys. Dept., Carleton Univ., Ottawa, Canada K1S 5B6}\\
E-mail: godfrey@physics.carleton.ca, kalyniak@physics.carleton.ca,
bkamal@physics.carleton.ca\\
$\nobody^b$ {\it
Ludwigs--Maximilians-Universit\"at, Sektion Physik, Theresienstr. 37,\\
D-80333 M\"unchen, Germany}\\
E-mail: leike@theorie.physik.uni-muenchen.de\\
\end{center}

\begin{abstract}
If extra gauge bosons are observed at the LHC or at a linear $e^+e^-$ collider,
the reaction $e^+e^-\rightarrow \nu\bar{\nu}\gamma$ can give 
information on the 
couplings $W'\nu l$ and $Z'\nu\bar\nu$ .
The total cross section and polarization asymmetries are the most sensitive 
observables for such a measurement.
Small systematic errors are crucial to obtain reasonable coupling constraints,
especially for a high luminosity collider.
\end{abstract}
%
\section{Introduction}
Extra charged ($W'$) and extra neutral ($Z'$) gauge bosons are predicted
by many extensions of the Standard Model (SM).
The masses and the couplings of the extra gauge bosons to fermions 
depend on the free parameters of the theory extending the Standard 
Model (SM).
They must be determined by experiment.
Hence, the observation of extra gauge bosons would teach us an important 
lesson on physics beyond the SM.
Therefore, the search for these particles is included in the research 
program of every future collider.

Recent collider data are consistent with the SM.
In particular, they suggest that extra gauge bosons predicted in 
Grand Unified Theories (GUT's) must be heavier than $\sim$500\,GeV \cite{PDB98}.
We therefore expect that extra gauge bosons would manifest themselves as 
(small) deviations of some observables from the SM predictions at future linear
$e^+e^-$ colliders.
At LHC, they would show up through peaks in the invariant mass distribution of 
their decay products.
Two cases are possible. 
If the SM is confirmed at future colliders, 
impressive improvements on the exclusion limits on extra gauge 
bosons can be achieved, see e.g. \cite{physrep,otherrefs} for a review and 
further references.
For extra gauge bosons with masses considerably smaller than their search 
limit in the same reaction, some information 
on their masses and couplings can be obtained.
See again \cite{physrep,otherrefs} for a review and further references.

In most models, the reactions $e^+e^-\rightarrow (\gamma,Z,Z')\rightarrow 
f\bar f$ and 
$pp\rightarrow Z'(W')X$ are most sensitive to a $Z'$ or $W'$. 
We suppose here that a signal of extra gauge bosons is detected in one of
these reactions and its mass and some of its couplings are measured.

In earlier contributions \cite{ourprocs}, we investigated the potential of 
the reaction $e^+e^-\rightarrow \nu\bar{\nu}\gamma$ in obtaining search limits
for a $W'$ appearing in different models. 
For some models, these limits can compete with those from the LHC.
In contrast to the limits from hadron colliders, the limits from $e^+e^-$ 
collisions are independent of unknown details of structure functions and
quark mixing.
This is a good process at an $e^+e^-$ collider for putting bounds on a $W'$.

In this contribution, we show where the reaction 
$e^+e^-\rightarrow \nu\bar{\nu}\gamma$ can contribute complementary information
to the coupling measurements of extra gauge bosons.
A more detailed paper is forthcoming.
%
\section{Assumptions and Calculation}
We show constraints on the couplings $L_\nu(Z'), R_\nu(Z')$ of extra neutral
gauge bosons to neutrinos and on the couplings $L_l(W'), R_l(W')$ 
of extra charged gauge bosons to leptons.
The couplings $L_f(V)$ and $R_f(V)$ of the left- and right-handed
fermions $f_{L,R}$ to the vector boson $V$ are defined by the following
interaction lagrangian,
\begin{equation}
{\cal L} = V_\mu\bar f_L\gamma^\mu f_L\cdot L_f(V)     
          +V_\mu\bar f_R\gamma^\mu f_R\cdot R_f(V).
\end{equation}

Two-dimensional constraints (95\% confidence) correspond to 
\begin{equation}
\label{chi2}
\chi^2=\sum_i\left(\frac{O_i(SM)-O_i(SM+Z'+W')}{\Delta O_i}\right)^2=5.99.
\end{equation}
In equation (\ref{chi2}), $O_i(SM)$ is the expectation for the observable 
$O_i$ by the SM, $O_i(SM+Z'+W')$ is the prediction of the 
extension of the SM and $\Delta O_i$ is the expected experimental error.
The index $i$ numbers different observables as, for example, the total
cross section $\sigma_T=\sigma$ and $A_{LR}$.

The physical input for our numerics is
$M_W = 80.33\,GeV, M_Z = 91.187 \,GeV, \sin^2\theta_W = 0.23124, 
\alpha=1/128$ and $\Gamma_Z=2.49 \,GeV$.

Let $E_\gamma$ and $\theta_\gamma$ denote the photon's energy and angle
in the $e^+e^-$ center-of-mass, respectively.
We have restricted
the range of $E_\gamma$ and $\theta_\gamma$ as follows:
\begin{equation}
10\,GeV < E_\gamma < \frac{\sqrt{s}}{2}(1-M_Z^2/s)-6\Gamma_Z,
\end{equation}
\begin{equation}
10^o < \theta_\gamma < 170^o,
\end{equation}
so that the photon may be detected cleanly. 
The upper bound on $E_\gamma$ ensures that the photons from the radiative
return to the SM $Z$ resonance are rejected.
As well, the angular cut eliminates the collinear singularity arising 
when the photon is emitted parallel to the beam.           

The most dangerous background to the reaction  
$e^+e^-\rightarrow \nu\bar{\nu}\gamma$ comes from 
radiative Bhabha scattering.
It can be eliminated by a cut on the transverse momentum of the photon,
\begin{equation}
P_{T,\gamma}>\sqrt{s}\sin\theta_\gamma\sin\theta_v/
(\sin\theta_\gamma + \sin\theta_v),
\end{equation}
where $\theta_v$ is the minimum angle for veto detectors.
We take $\theta_v=25$\,mrad.

The unpolarized cross section is composed of polarized cross sections as
$\sigma=\sigma_{LL} + \sigma_{LR} + \sigma_{RL} + \sigma_{RR} 
= \sigma_L  + \sigma_R$.
Note that $\sigma_{LL}=\sigma_{RR}\ne 0$ only in models where the $W'$ has 
non-zero couplings to left- and right-handed fermions simultaneously.
Even in this case, we have  
$\sigma_{LL},\sigma_{RR} \ll [\sigma(SM)-\sigma(SM+Z'+W')]$ due to present
experimental bounds on extra gauge bosons and because 
$\sigma_{LL}$ and $\sigma_{RR}$ do not interfere with the SM.

We consider polarization asymmetries involving polarized electrons only, 
$P^-=90\%$, $A_{LR}= (\sigma_L - \sigma_R)/(\sigma_L  + \sigma_R)$, 
and polarization asymmetries involving both polarized beams, 
$P^-=90\%$, $P^+=60\%$, 
$A_{LR}= (\sigma_{LR} - \sigma_{RL})/(\sigma_{LR}  + \sigma_{RL})$ and devote 
half of the luminosity to each polarization combination. 
The case of two polarized beams corresponds to an effective polarization
$P_{eff}=(P^-+P^+)/(1+P^-P^+)=97.4\%$ \cite{marciano} of the electron beam.
Note that the use of the effective polarization is exact for 
$\sigma_{LL}=\sigma_{RR} = 0$. 
If this is not the case, it remains a good approximation.

We assume systematic errors of 1\% for both observables $\sigma$ and $A_{LR}$,
although we know that this assumption is very optimistic for $\sigma$.

We have performed two independent calculations.
In the first calculation, the squared matrix element was obtained in the
conventional trace formalism. The phase space of the neutrinos was
integrated out analytically. This results in (rather long) analytical
formulas for the doubly differential cross section
$d^2\sigma/(dE_\gamma d\cos\theta_\gamma)$. All analytical integrals
were checked numerically to a precision of at least 10 digits.
Artificial singularities in the physical phase space appeared from the
partial fraction decomposition of the first integration. 
These singularities were proven to cancel analytically after the
second integration and an appropriate Taylor expansion. 
All algebraic manipulations were carried out with the algebraic
manipulation program {\tt form} \cite{form}.
The remaining one (two) integrations to obtain simple distributions
(total cross sections) were done using an adaptive Simpson integration
routine.

In a second calculation, the helicity amplitudes were calculated by spinor techniques.
They were squared numerically and summed over the helicities of the 
final states.
The result was integrated using Monte Carlo techniques.

The total cross sections from both calculations agree 
within the Monte Carlo errors.
For the SM, they agree with the predictions of {\tt COMPHEP} \cite{comphep}.
%
\section{Extra Neutral Gauge Bosons}
\begin{figure}[tbh]
\ \vspace{1cm}\\
\begin{minipage}[t]{7.8cm} {
\begin{center}
\hspace{-1.7cm}
\mbox{
\epsfysize=7.0cm
\epsffile[0 0 500 500]{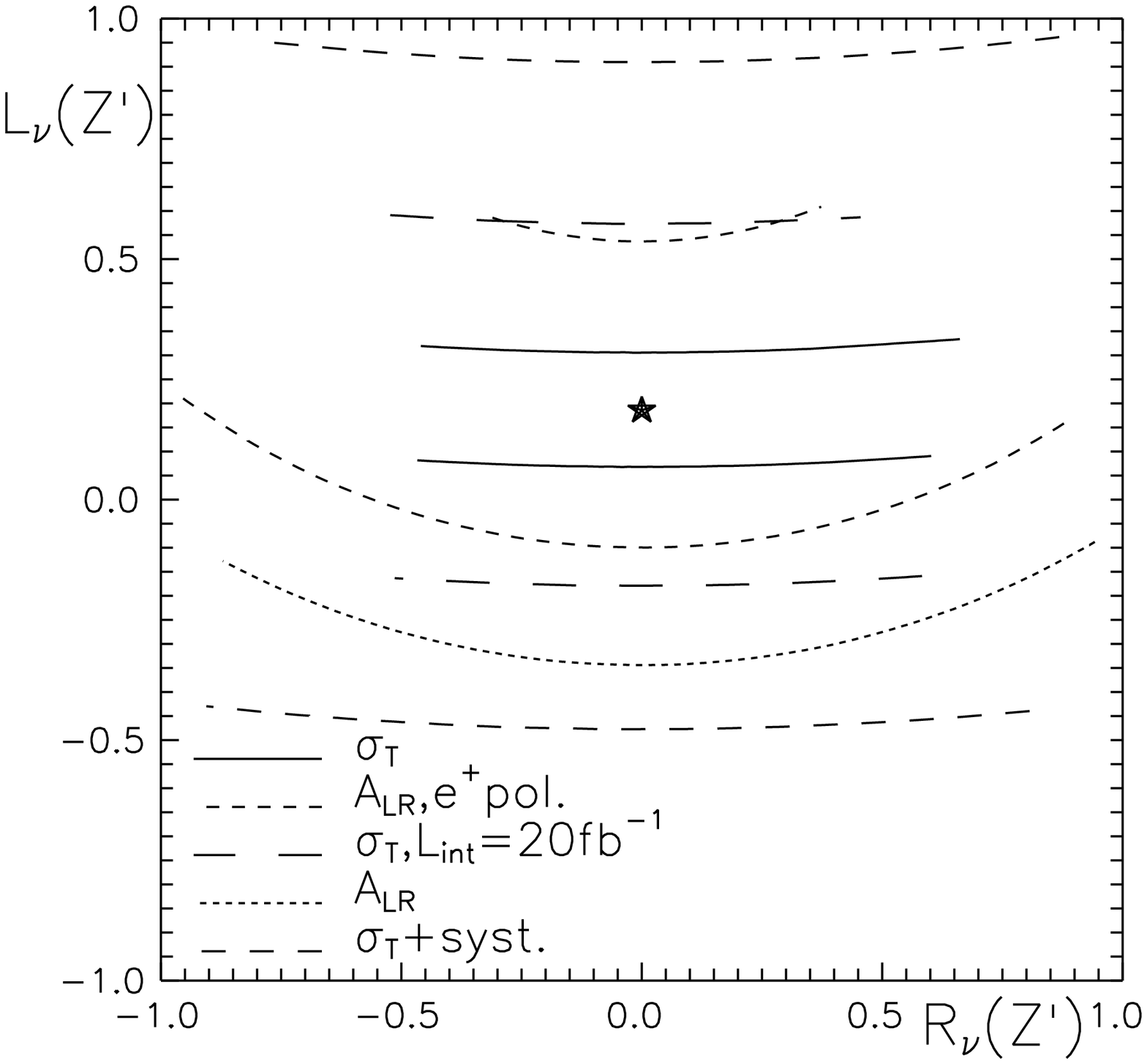}
}
\end{center}
\vspace*{-0.5cm}
\noindent
{\small\it
\begin{fig} \label{nng1} 
Constraints on the $Z'\nu\bar\nu$ couplings $R_\nu(Z')$ and $L_\nu(Z')$ 
by different observables,see text.
$\sqrt{s}=0.5\,TeV,\ M_{Z'}=1.5\,TeV$.
$L_{int}=200\,fb^{-1}$ except the indicated case were it is $20\,fb^{-1}$.
The polarization of the electron beam is 90\%.
The polarization of the positron beam (where indicated) is 60\%.
Where indicated, the systematic error is 1\% for $\sigma$ and $A_{LR}$, 
otherwise it is zero .
The assumed model is a $Z'$ in the Sequential Standard Model (SSM),
indicated by a star.
\end{fig}}
}\end{minipage}
\hspace*{0.5cm}
\begin{minipage}[t]{7.8cm} {
\begin{center}
\hspace{-1.7cm}
\mbox{
\epsfysize=7.0cm
\epsffile[0 0 500 500]{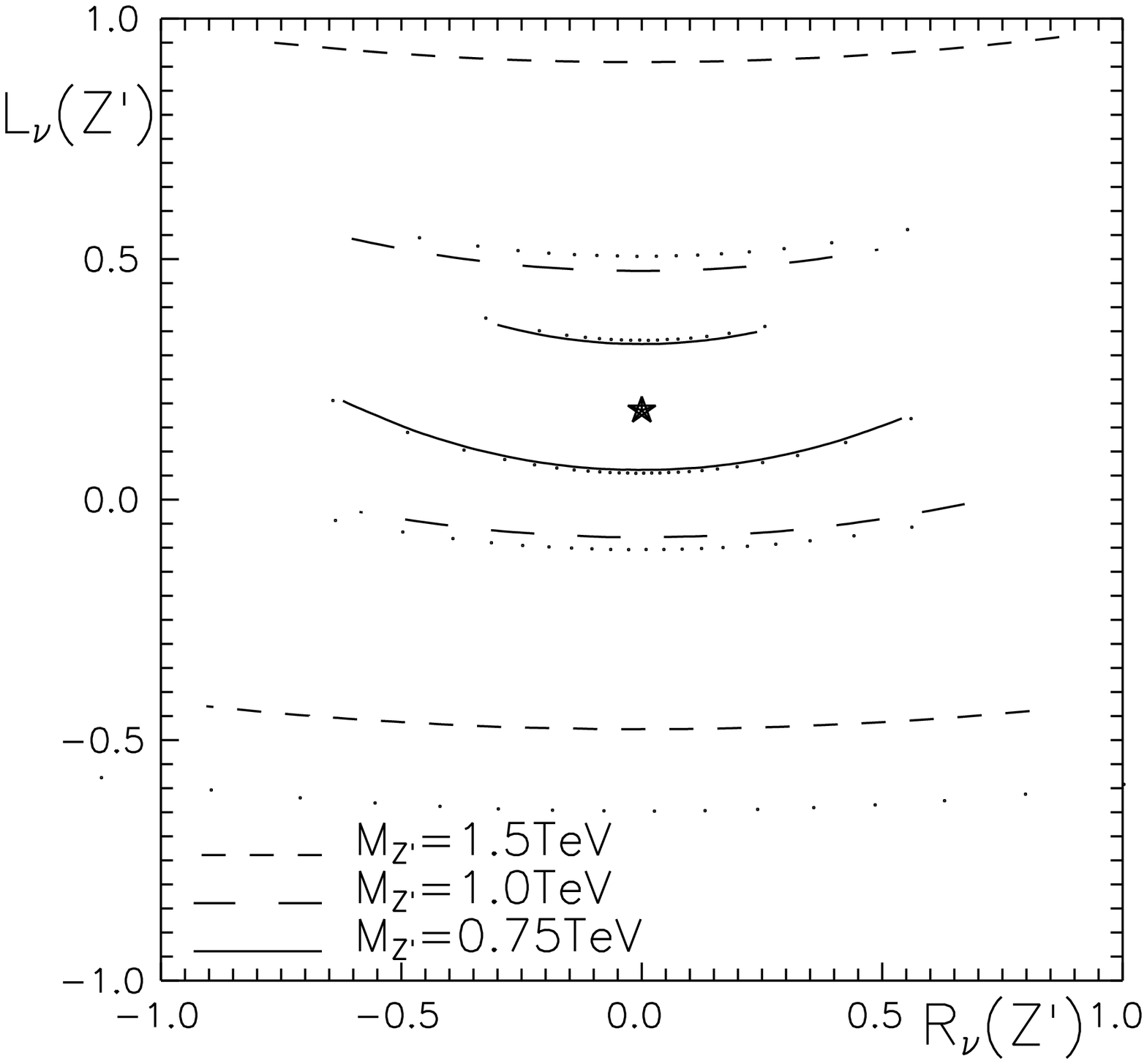}
}
\end{center}
\vspace*{-0.5cm}
\noindent
{\small\it
\begin{fig} \label{nng2} 
Constraints on  $R_\nu(Z')$ and $L_\nu(Z')$ below the $Z'$ peak by $\sigma$
only.
The lines show the results for three different $Z'$ masses.
The dots indicate how the constraints relax if the error of the $Z'e^+e^-$ 
coupling measurement is included. $\sqrt{s}=0.5\,TeV$,\ $L_{int}=200\,fb^{-1}$ 
A systematic error of 1\% is included.
The assumed model (SSM $Z'$) is indicated by a star.
\end{fig}}
}\end{minipage}
\end{figure}

In figures~\ref{nng1} and \ref{nng2}, 
we assume that there is a signal from a $Z'$ and no signal from a $W'$.
This can happen in models where the $W'$ has purely right-handed
couplings and the right-handed neutrino is heavy.
If there is a signal for a $W'$, a similar analysis could be done including
the known $W'$ parameters. The experimental errors of these parameters would 
then enlarge the errors of the measurements shown in the next two figures.
However, it would leave our main conclusions unchanged.

We find that the process $e^+e^-\rightarrow\nu\bar\nu\gamma$ 
can give some constraints to the couplings of 
the $Z'$ to SM neutrinos {\it below} the $Z'$ resonance.

In figures~\ref{nng1} and~\ref{nng2}, we assume a heavy repetition of the 
SM Z boson. 
We see that we can get some constraints even in the 
case where the $Z'$ is considerably heavier than the center-of-mass energy.
The region which cannot be resolved by the observables is between the 
two lines and contains the couplings of the assumed model.
This choice also shows our normalization of the couplings.
For the cases where only one bounding line is shown, the second line
is outside the figure.
$R_\nu(Z')$ and $L_\nu(Z')$ are mainly constrained by the interference of the 
$Z'$ contributions with the SM contributions.
Mainly the $Z'$ coupling to left-handed neutrinos can be constrained.
This makes the constraints especially simple.

Figure~\ref{nng1} illustrates the effect of different observables and 
different experimental parameters on these constraints.
The total cross section gives the strongest constraint.
The constraint from $A_{LR}$ is shown for one and two polarized beams.
The constraints from energy and angular distributions with 
10 equal size bins give no improvement.
The constraint for a low luminosity linear collider 
is also shown in figure~\ref{nng1}.
A systematic error of 1\% relaxes the constraints considerably and 
dilutes the advantage of high luminosity.
We see that small systematic errors and a high luminosity collider 
are highly desired for the proposed model measurement.

Figure~\ref{nng2} shows the possible constraints on  $R_\nu(Z')$ and 
$L_\nu(Z')$ 
including a systematic error of 1\% for three representative 
$Z'$ masses.
The constraints become much stronger for smaller $Z'$ masses.
So far, we assumed that the $Z'e^+e^-$ couplings  $R_e(Z')$ and $L_e(Z')$ 
are precisely known.
However, they must be measured (with errors) by experiment.
Figure~4b of reference \cite{lmu0296} illustrates such a measurement 
for a collider with conventional luminosity. 
The errors from reference \cite{lmu0296} are taken to estimate its influence on the
$R_\nu(Z')$, $L_\nu(Z')$ constraint. 
Our input for the errors of the $Z'e^+e^-$ couplings for $M_{Z'}=1.0\,TeV$ and
$0.75\,TeV$ are obtained from those for $1.5\,TeV$ by the scaling relation
(2.63) in \cite{physrep}.
We see that 
the uncertain knowledge of the $Z'e^+e^-$ couplings leads to weaker constraints
on $R_\nu(Z')$ and $L_\nu(Z')$. 
However, figure~\ref{nng2} shows that this effect is only important 
for a very heavy
$Z'$ where its couplings are already weakly constrained.
The influence of the errors of the $Z'e^+e^-$ coupling measurement
is negligible for a relatively light $Z'$.

Finally, we mention that there is no sign ambiguity in the measurement 
of  $R_\nu(Z')$ and $L_\nu(Z')$, if the signs of the  $Z'e^+e^-$ couplings 
are known.
Remember \cite{lmu0296} that the $Z'e^+e^-$ couplings have a two-fold sign
ambiguity if measured in the reaction $e^+e^-\rightarrow f\bar f$ alone.
If the sign ambiguity in the $Z'e^+e^-$ couplings is removed \cite{physrep}, 
i.e. by measurements of the reaction  $e^+e^-\rightarrow W^+W^-$ below the
$Z'$ resonance or by measurements at the $Z'$ resonance, it disappears also 
in the proposed constraint of $R_\nu(Z')$ and $L_\nu(Z')$.
%
\section{Extra Charged Gauge Bosons}
\begin{figure}[tbh]
\ \vspace{1cm}\\
\begin{minipage}[t]{7.8cm} {
\begin{center}
\hspace{-1.7cm}
\mbox{
\epsfysize=7.0cm
\epsffile[0 0 500 500]{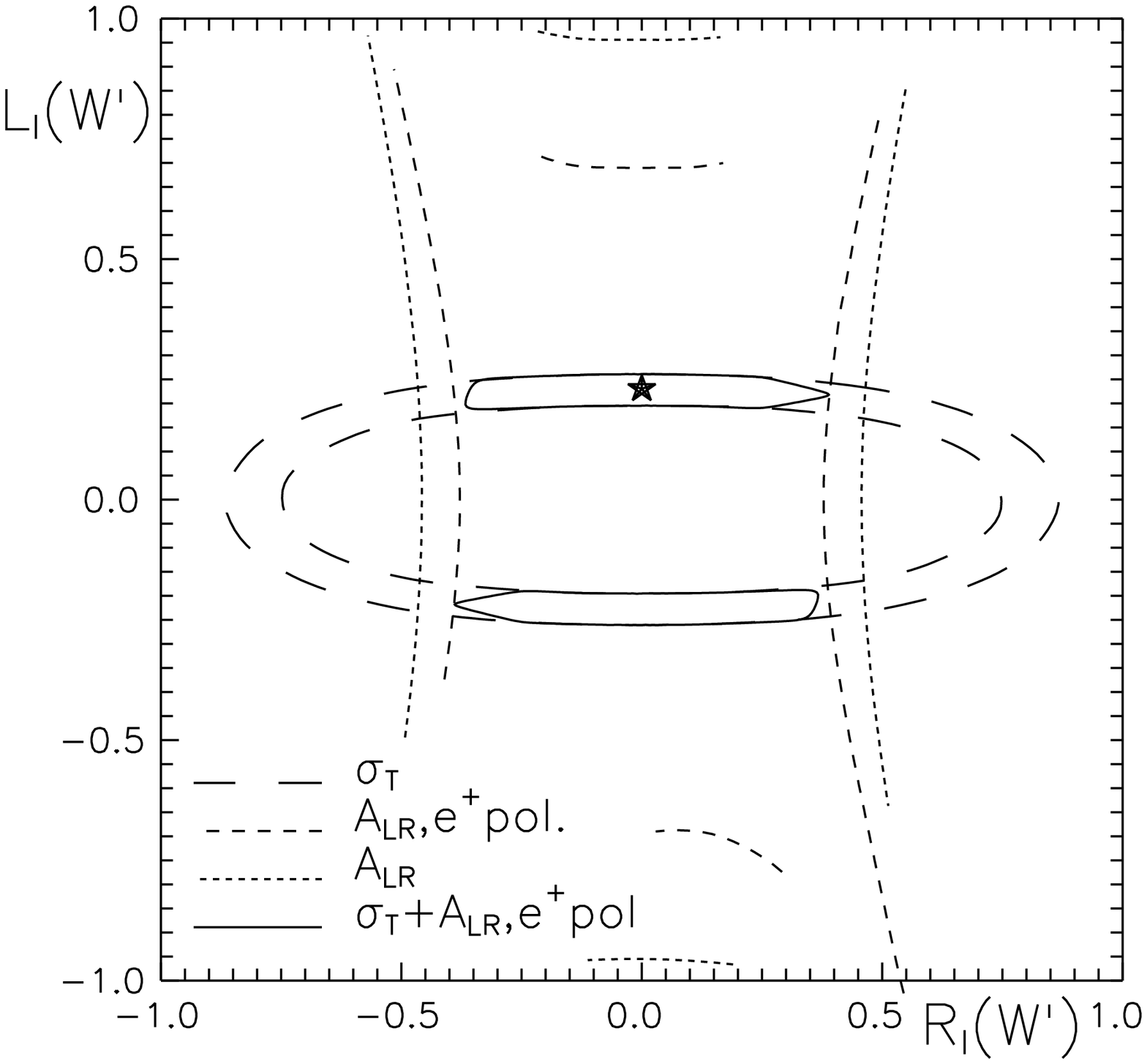}
}
\end{center}
\vspace*{-0.5cm}
\noindent
{\small\it
\begin{fig} \label{nng3} 
Constraints on the $W'$ couplings by $\sigma$ and $A_{LR}$.
$\sqrt{s}=0.5\,TeV,\ L_{int}=200\,fb^{-1}$ and $M_{W'}=1.5\,TeV$.
The constraint by $\sigma$ and $A_{LR}$ combined is also shown.
Only statistical errors are included in this figure.
90\% electron and, where indicated, 60\% positron polarization are assumed.
The assumed model (SSM $W'$) is indicated by a star.
\end{fig}}
}\end{minipage}
\hspace*{0.5cm}
\begin{minipage}[t]{7.8cm} {
\begin{center}
\hspace{-1.7cm}
\mbox{
\epsfysize=7.0cm
\epsffile[0 0 500 500]{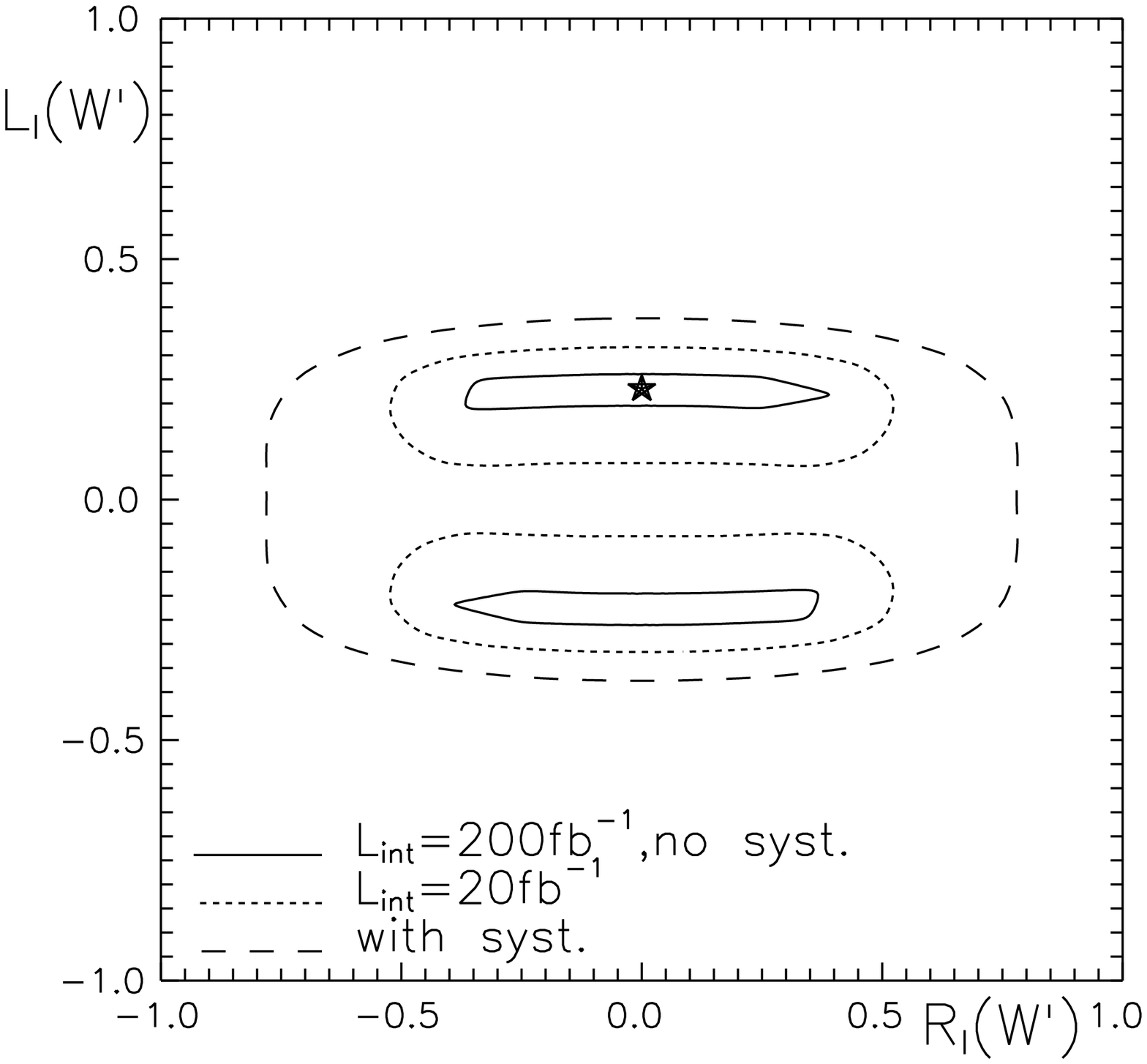}
}
\end{center}
\vspace*{-0.5cm}
\noindent
{\small\it
\begin{fig} \label{nng4} 
Constraints on the $W'$ couplings by $\sigma$ and $A_{LR}$ combined.
90\% electron and 60\% positron polarization are assumed. 
$\sqrt{s}=0.5\,TeV,\ M_{W'}=1.5\,TeV$. 
$L_{int}=200\,fb^{-1}$ except the indicated case where it is $20\,fb^{-1}$.
Only statistical errors are included except the indicated case where a 
systematic error of 1\% is included for $\sigma$ and $A_{LR}$.
The assumed model (SSM $W'$) is indicated by a star.
\end{fig}}
}\end{minipage}
\end{figure}

In figures~\ref{nng3} to 6, we assume that there is no signal from a 
$Z'$ but a signal from a $W'$.
This can happen in models where the $W'$ is considerably 
lighter than the $Z'$.
We recognize that this particular scenario is unlikely in the context 
of the models we consider. 
For instance, in the UUM model, the W and Z masses are approximately equal 
and there would most likely be a signal observed for the $Z'$ in addition to 
the $W'$.
So it should be understood that our results for the case of a $W'$ only
represent an estimate of the reach of this process in constraining $W'$
couplings, rather than precision limits in the context of a fuller 
understanding of the physics realized in nature. We use this simple scenario
in order to indicate sensitivity to various parameters, such as the
observables used and the luminosity.
In the case of a $Z'$ signal, the knowledge about this particle should
be included in the following analysis.
Again, the experimental errors of the measurements of the $Z'$ parameters
would enlarge the errors of the $W'$ measurements but not change our
main conclusions.

We find that the process $e^+e^-\rightarrow\nu\bar\nu\gamma$ can give 
constraints on the couplings $L_l(W')$ and $R_l(W')$ 
for $W'$ masses considerably larger than the center-of-mass energy.

Figure~\ref{nng3} is similar to figure~\ref{nng1} but 
it shows the constraint on the $W'$ couplings.
For illustration, we assume that there exists a $W'$ with 
couplings as the SM W but with a mass of $1.5\,TeV$.
We get some constraints on the couplings even in this case.
The constraints from energy and angular distributions give 
no improvement for the considered model.
The constraint from $A_{LR}$ is complementary to that from the total cross 
section.
It is shown for one and two polarized beams.
It turns out that $\sigma$ and $A_{LR}$ together give the 
best constraint on the coupling.
The constraints on the $W'$ couplings always have a two-fold sign ambiguity,
i.e. nothing is changed by a simultaneous change of the sign of 
$L_l(W')$ and $R_l(W')$.
This ambiguity is obvious from the amplitude where $W'$ couplings
always enter in products of pairs.

In figure~\ref{nng4}, we show constraints on the $W'$ couplings from $\sigma$ and 
$A_{LR}$ combined.
We have two well separated 
regions for high luminosity and no systematic error
(compare figure~\ref{nng3}).
These two regions come closer together for low luminosity and 
no systematic error.
We are left with one large region after the inclusion of a systematic error 
of 1\%.
As in the case of extra neutral gauge bosons, 
a small systematic error
{\it and} a high luminosity are necessary for a coupling measurement.

\begin{figure}[tbh]
\ \vspace{1cm}\\
\begin{minipage}[t]{7.8cm} {
\begin{center}
\hspace{-1.7cm}
\mbox{
\epsfysize=7.0cm
\epsffile[0 0 500 500]{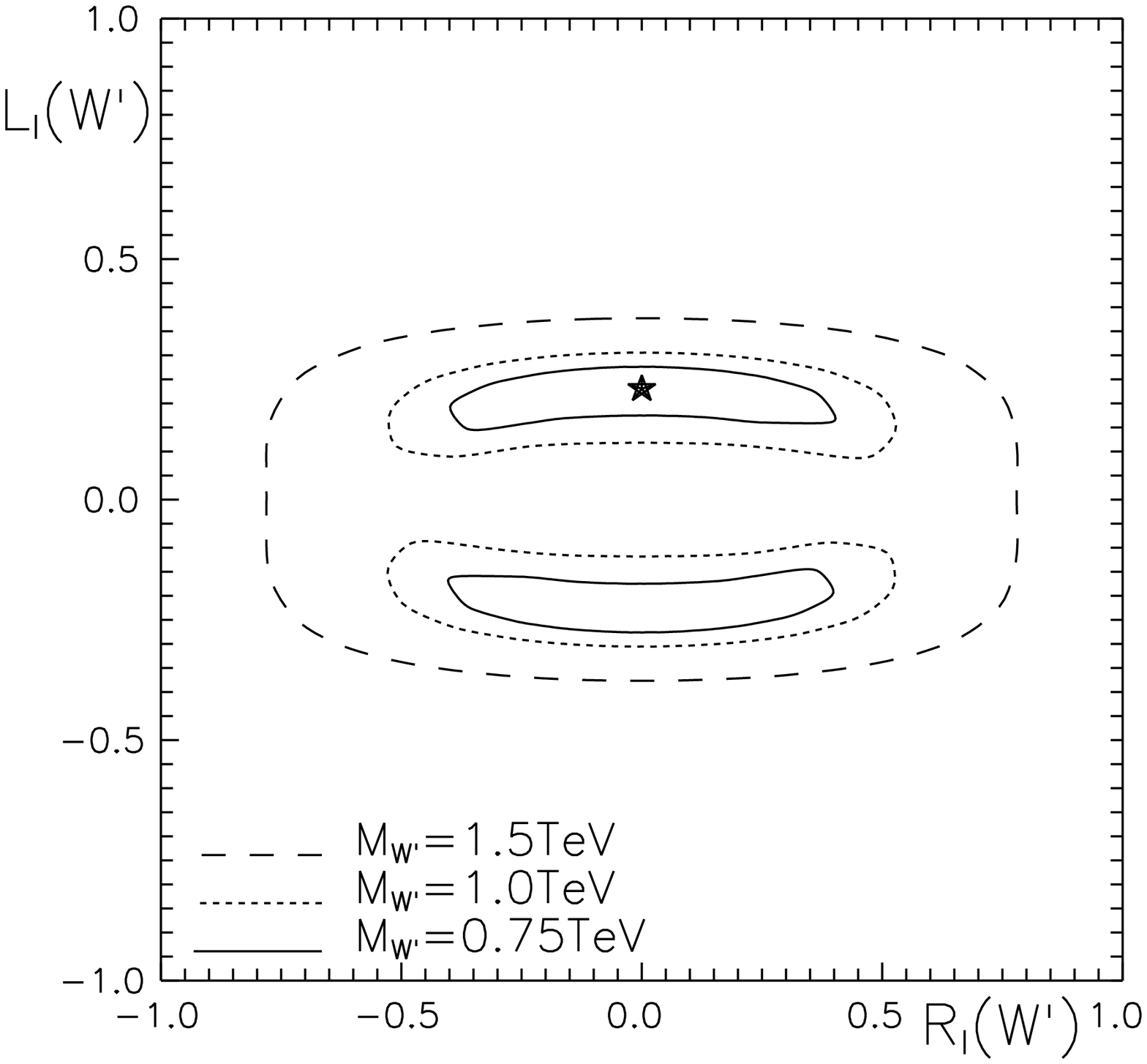}
}
\end{center}
\vspace*{-0.5cm}
\noindent
{\small\it
\begin{fig} \label{nng5} 
Constraints on the $W'$ couplings by $\sigma$ and $A_{LR}$ combined
for different $W'$ masses.
90\% electron and 60\% positron polarization,
$\sqrt{s}=0.5\,TeV$ and $L_{int}=200\,fb^{-1}$ are assumed.
A systematic error of 1\% is included for $\sigma$ and $A_{LR}$.
The assumed model (SSM $W'$) is indicated by a star.
\end{fig}}
}\end{minipage}
\hspace*{0.5cm}
\begin{minipage}[t]{7.8cm} {
\begin{center}
\hspace{-1.7cm}
\mbox{
\epsfysize=7.0cm
\epsffile[0 0 500 500]{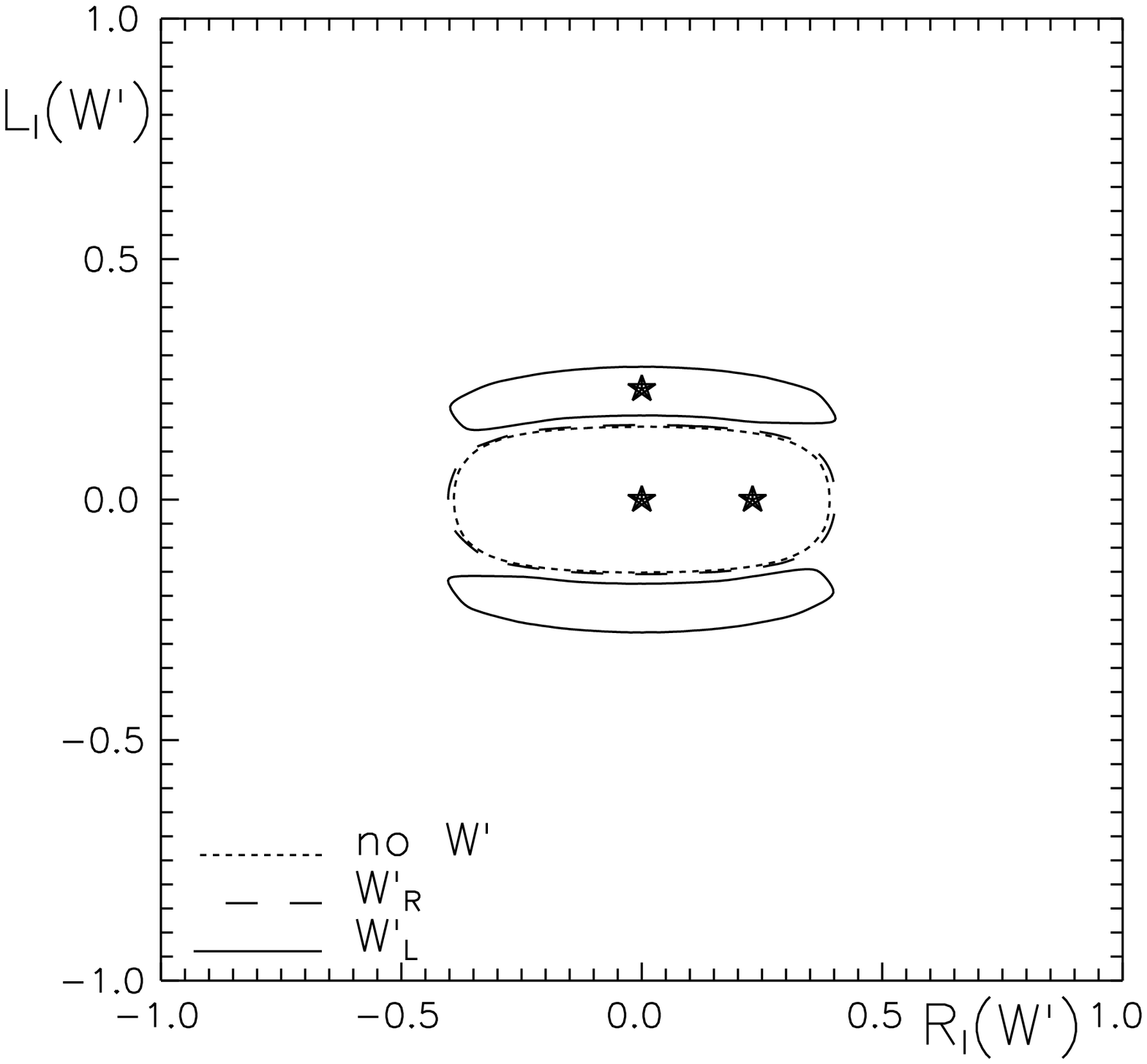}
}
\end{center}
\vspace*{-0.5cm}
\noindent
{\small\it
\begin{fig} \label{nng6} 
Constraints on the $W'$ couplings by $\sigma$ and $A_{LR}$ combined
for different $W'$ scenarios.
90\% electron and 60\% positron polarization,
$\sqrt{s}=0.5\,TeV$, $L_{int}=200\,fb^{-1}$ and $M_{W'}=0.75\,TeV$ are 
assumed. 
A systematic error of 1\% is included for $\sigma$ and $A_{LR}$.
The assumed models are indicated by a star.
\end{fig}}
}\end{minipage}
\end{figure}

In figure~\ref{nng5}, we show how the constraints on the $W'$ couplings vary 
for different $W'$ masses.
The constraint for $M_{W'}=1.5\,TeV$ is identical to that shown in 
figure~\ref{nng4}.
We see that the constraint on the $W'$ masses improves dramatically 
for lower $W'$ masses.

Figure~\ref{nng6} illustrates the possible 
discrimination between different models.
We see that a $W'$ with SM couplings ($W'_L$) can be separated from the SM.
A $W'$ with pure right-handed couplings ($W'_R$) with a strength of the 
left-handed coupling of the SM $W$ cannot be distinguished from the SM case. 

From the amplitude of  the process $e^+e^-\rightarrow\nu\bar\nu\gamma$,
it can be seen
that the constraints shown in figs.~\ref{nng3} to \ref{nng6} are,
to good approximation, 
valid for the combinations $L_l(W')/M_{W'}$ and  $R_l(W')/M_{W'}$, and not
for the couplings and the mass separately.
We have fixed so far the $W'$ mass for illustration.
If a $W'$ is found with a mass different from our assumptions, the constraint
on its couplings can then be obtained by the appropriate scaling of our 
results as long as $M_{W'}\gg\sqrt{s}$.

\begin{figure}[tbh]
\ \vspace{1cm}\\
\begin{minipage}[t]{7.8cm} {
\begin{center}
\hspace{-1.7cm}
\mbox{
\epsfysize=7.0cm
\epsffile[0 0 500 500]{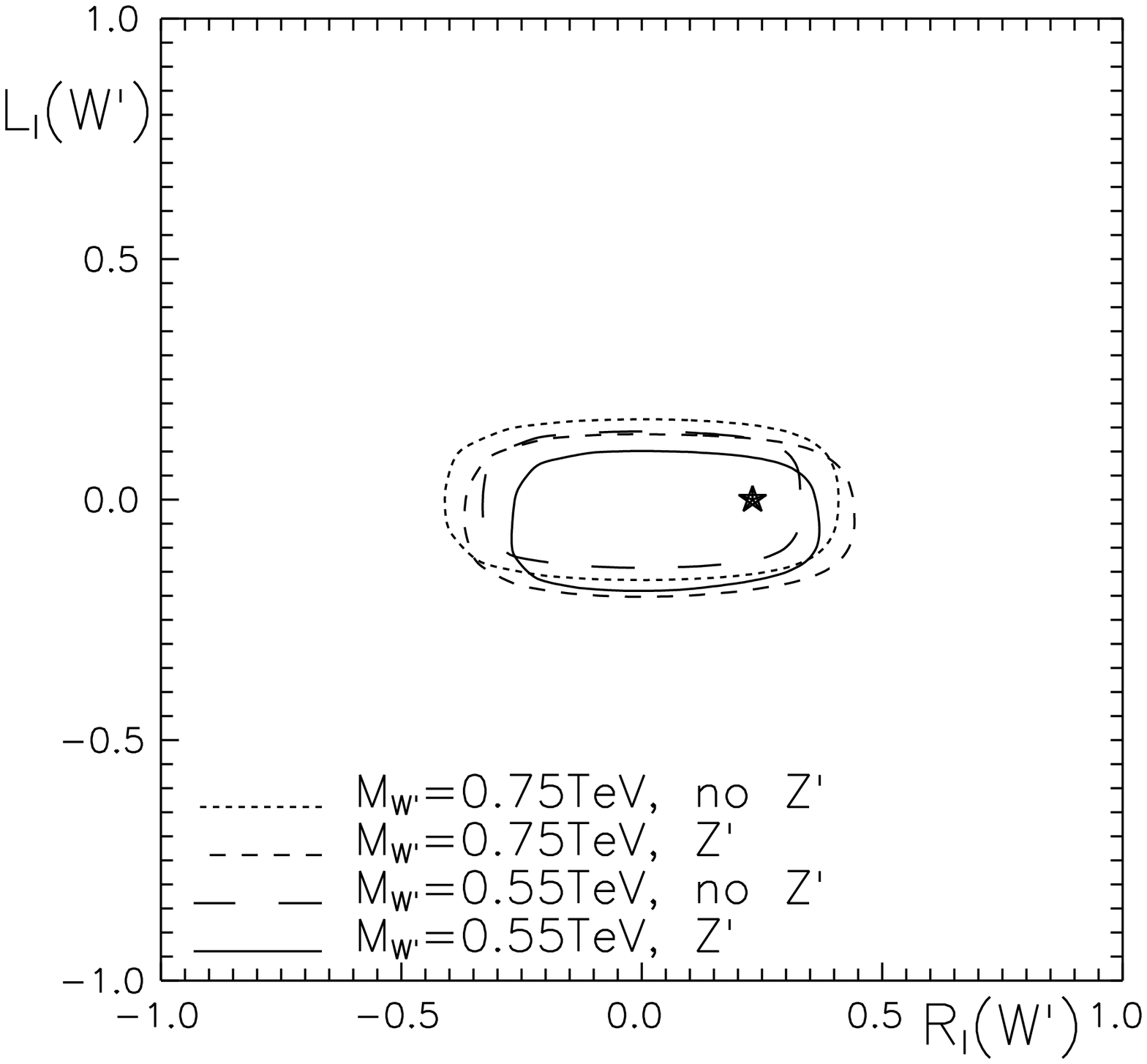}
}
\end{center}
\vspace*{-0.5cm}
\noindent
{\small\it
\begin{fig} \label{nng7} 
Constraints on the $W'$ couplings by $\sigma$ and $A_{LR}$ combined
in the LR model with $\rho=1$ and $\kappa=1$
for different $W'$ masses and different fitting strategies (see text).
90\% electron and 60\% positron polarization,
$\sqrt{s}=0.5\,TeV$ and $L_{int}=200\,fb^{-1}$ are assumed.
A systematic error of 1\% is included for $\sigma$ and $A_{LR}$.
The assumed model is indicated by a star.
\end{fig}}
}\end{minipage}
\hspace*{0.5cm}
\begin{minipage}[t]{7.8cm} {
\begin{center}
\hspace{-1.7cm}
\mbox{
\epsfysize=7.0cm
\epsffile[0 0 500 500]{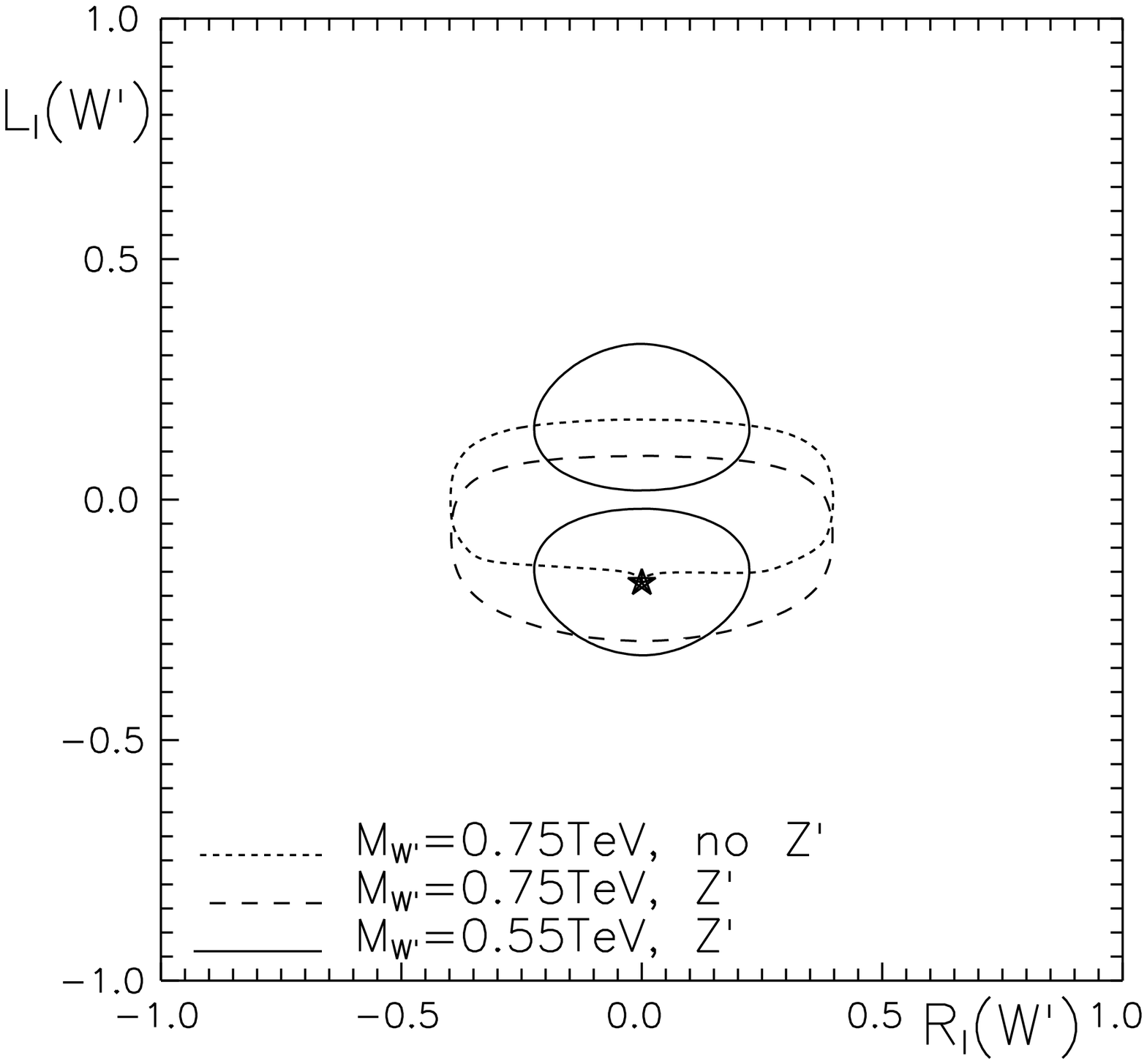}
}
\end{center}
\vspace*{-0.5cm}
\noindent
{\small\it
\begin{fig} \label{nng8} 
Constraints on the $W'$ couplings by $\sigma$ and $A_{LR}$ combined
in the UUM with $\sin\psi=0.6$
for different $W'$ masses and different fitting strategies (see text).
90\% electron and 60\% positron polarization,
$\sqrt{s}=0.5\,TeV$ and $L_{int}=200\,fb^{-1}$ are assumed.
A systematic error of 1\% is included for $\sigma$ and $A_{LR}$.
The assumed model is indicated by a star.
\end{fig}}
}\end{minipage}
\end{figure}

We considered model independent bounds on the couplings of extra
gauge bosons neglecting the existence of other extra gauge bosons, i.e.
of the $W'$ in the case of $Z'$ constraints or of the $Z'$ in the case of 
$W'$ constraints.
However, in the general case, extra neutral and extra charged gauge bosons 
simultaneously influence the observables.

In figure~\ref{nng7}, we assume that the left-right-symmetric model is true.
See reference \cite{ourprocs} for conventions.
In the first case, we take $M_{W'}=0.75\,TeV$. 
It follows that $M_{Z'}=0.90(1.27)\,TeV$ for $\kappa=1$ and $\rho=1(2)$.
We show the constraints on the couplings of the $W'$ for $\rho=1$ 
obtained by two different fitting strategies:
first ignoring the $Z'$ completely, and second taking the $Z'$ into account
exactly without any errors.
We see that the two curves are quite close.
The reason is that the reaction  $e^+e^-\rightarrow\nu\bar\nu\gamma$ 
is not very sensitive to such a $Z'$.

The more realistic case that the $Z'$ parameters are measured with some errors
would 
produce a constraint between the two curves of figure~\ref{nng7}. 
The case $\rho=2$ predicts a heavier $Z'$.
This would produce two curves, which differ even less from each other 
than those for $\rho=1$. 
To demonstrate how the constraints change for a larger signal,
we repeated the same procedure 
with $M_{W'}=550\,GeV$. This number (and the mass of the associated $Z'$)
are at the edge of the present exclusion limit \cite{PDB98}.
However, the constraints get only a minor improvement.

Figure~\ref{nng8} is similar to figure~\ref{nng7}, 
but now we assume that the Un-Unified model is true.
See again reference \cite{ourprocs} for conventions.
We consider two different mass scenarios, $M_{W'}=M_{Z'}=0.75\,TeV$ and
$M_{W'}=M_{Z'}=0.55\,TeV$.
We show the constraints on the couplings of the $W'$ obtained by the same 
two fitting strategies as demonstrated in figure~\ref{nng7}.
Already for masses of $0.75\,TeV$, the two curves differ much more than 
was the case for the LR model.
For masses of $0.55\,TeV$, the wrong fitting strategy gives always a
$\chi^2$ above 15, i.e. no allowed region in our case.
This shows that such a light $Z'$ cannot be 
ignored in the fitting procedure.

Other reactions as $e^+e^-\rightarrow f\bar f$ or hadron collisions are
more sensitive to a $Z'$ than $e^+e^-\rightarrow\nu\bar\nu\gamma$.
In almost all cases, they will detect a $Z'$ signal in the cases where 
the $Z'$ contribution
is relevant for a $W'$ constraint by $e^+e^-\rightarrow\nu\bar\nu\gamma$.
We conclude from figures \ref{nng7} and \ref{nng8} that this information
from other experiments is needed for a reliable  constraint on the
$W'$ coupling.
%
\section{Conclusions}
We found that the reaction $e^+e^-\rightarrow \nu\bar{\nu}\gamma$ can give 
interesting constraints on the couplings of extra gauge bosons to leptons.
The total cross section and polarization asymmetries are the most 
sensitive observables for this measurement.
The assumed universal systematic error of 1\% seems to be very optimistic
for the total cross section. 
Already this small systematic error almost dilutes the advantages gained from
the high luminosity option of the collider.
A polarized electron beam is crucial. 
A polarized positron beam is not essential but gives a quantitative 
improvement of the coupling constraints.

\end{document}